\documentclass[conference]{IEEEtran}
\IEEEoverridecommandlockouts
\usepackage[font=small]{caption}
\usepackage{graphicx}
\graphicspath{ {./Pics/} }
\usepackage{amsmath}

\usepackage{cite}

\ifCLASSINFOpdf
\else
\fi
\usepackage{array}
\usepackage{url}

\usepackage{multicol}
\usepackage{caption}
\usepackage{subcaption}

\usepackage{flushend}

\usepackage{xcolor}
\begin{document}
%
\title{An Incremental Negative Sequence Admittance Method for Fault Detection in Inverter-Based Microgrids}

\author{\IEEEauthorblockN{Kwasi Opoku, Subash Pokharel, and Aleksandar Dimitrovski}
\IEEEauthorblockA{\textit{University of Central Florida}\\
Orlando, Florida\\}
}


%


\maketitle

\begin{abstract}
Superimposed sequence quantities have been relied on for years to provide pure fault components for various applications in protection schemes. In more recent times, they have been employed in various solutions to the challenges introduced by integration of distributed generation in distribution systems. 

To improve the overall reliability of protection schemes deployed in such active distribution networks, this paper evaluates a new approach for detecting unbalanced faults using negative sequence quantities. A proposed superimposed negative sequence admittance element is validated using simulation tests on a modelled community microgrid. The performance of the method is tested for various fault scenarios in MATLAB/Simulink. 
\end{abstract}


%
\IEEEpeerreviewmaketitle

\section{Introduction}
The challenges to fault detection in microgrids have been extensively discussed in various literature \cite{miveh2012review,memon2015critical,beheshtaein2019review}. Many of challenges arise from the bidirectional flow of fault currents in active distribution networks. This is a departure from conventional radial distribution networks which have been the basis of many existing protection schemes. Thus, old schemes deployed in active distribution networks are at risk of sympathetic tripping and protection blinding. Another major setback, particularly in the case of inverter-based distributed generators (IBDGs), is the thermal limitation imposed by the power electronic components of the inverters. In order to protect these components from damage under fault conditions, it is required for the steady state fault current contribution of such IIDGs to be limited to 1.2pu \cite{8332112}. In this case, protection schemes employed in distribution networks require novel detection methods that can effectively distinguish between load and fault conditions in such inverter-based microgrids. Other contributory factors to the protection challenges are the varying operation modes and topologies of active distribution networks, and the different fault current characteristics of the different types of distributed generators (DGs).

Many solutions have been developed by engineers and researchers to microgrid protection issues. A common approach is to employ adaptive protection methods that modify settings based on prevailing network topology and system values \cite{6758374,9292082}. This approach may be centralized, in which case, a central controller communicates new settings or setting change decision for instance via IEC 61850 IEC 60870–5-104 based communication as deployed in \cite{6281478}. Decentralized methods rely on peer-to-peer communication between field relays to determine appropriate settings. Clearly, the specific implementation of adaptive protection varies with each system, and as such intricate networks require extensive considerations of the various topology scenarios.

Some solutions employ methods used mainly in transmission networks. Due to the bidirectional nature of fault current flow in transmission networks, such protection philosophies may be adapted for active distribution systems. For instance, differential and distance protection applications for microgrids was proposed in \cite{6695477} and \cite{7392186} respectively. However, the presence of lateral feeders in distribution networks is a setback to the effectiveness of differential schemes. Similarly, the success of distance protection deployed in distribution networks is adversely affected by the short line lengths and the prevalence of high impedance faults.

Another approach to dealing with microgrid protection challenges is to develop or apply new quantities for detection of fault incidence that do not rely on high fault current magnitude. Travelling wave (TW) based protection schemes achieve fast fault detection by using fault-caused electromagnetic waves that propagate at the speed of light. This was implemented in \cite{8086193} for microgrid protection. Similar to distance and differential protection schemes, TW applications are effective for long lines, and in lines with fewer laterals. Superimposed sequence components, as used in high speed transmission protection applications \cite{benmouyal1999superimposed,sirisha2014incremental}, has been variously applied in microgrid protection to achieve the desired sensitivity, selectivity and dependability. These are incremental quantities that provide pure fault quantities that are immune to the influence of normal loading in the system. \cite{el2017fault, 9750756}. The theory and applications of the superimposed quantities are presented in the next section.

This paper evaluates a new approach for detecting unbalanced faults using superimposed sequence quantites. A proposed negative sequence admittance element is validated using simulations of fault cases in a community microgrid. The theory of superimposed quantities and their application in microgrid protection are discussed in Section II. The superimposed negative sequence admittance method and the test results is discussed in Sections III and IV respectively. Section V concludes the paper.

\section{A Review of Superimposed Sequence Quantities}
\subsection{Fundamentals of Superimposed Quantities}
Superimposed quantities are obtained by comparing quantities in the faulted and pre-fault state. Using the theory of superposition, a fault-driven circuit can be obtained whose quantities are pure fault quantities, not influenced by load.

\begin{figure}[!t]
\centering
\includegraphics[clip, trim=0.5cm 15.5cm 0.5cm 0cm, width=0.48\textwidth]{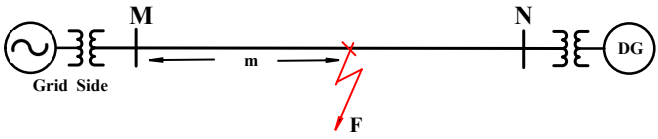}
\caption{Sample Network}
\label{fig_network}
\end{figure}

\begin{figure}[ht]
    \centering
    \includegraphics[clip, trim=0.5cm 13cm 0.2cm 0cm, width=0.45\textwidth]{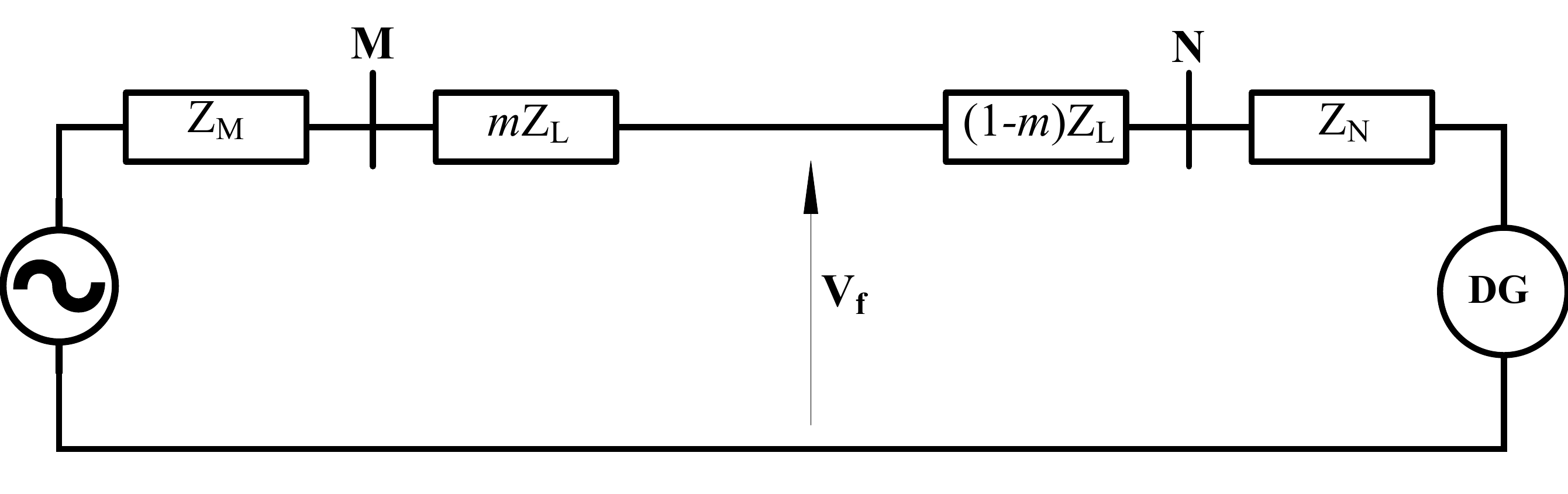}
    \caption{Pre-fault Network}
    \label{Prefault}
\end{figure}

\begin{figure}[ht]
    \centering
    \includegraphics[clip, trim=0.5cm 13cm 0.18cm 0cm, width=0.45\textwidth]{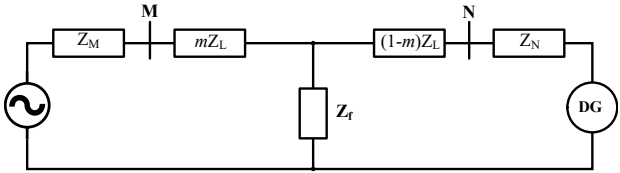}
    \caption{Faulted Network}
    \label{Fault}
\end{figure}

\begin{figure}[ht]
    \centering
    \includegraphics[clip, trim=0.5cm 12cm 0.2cm 0cm, width=0.45\textwidth]{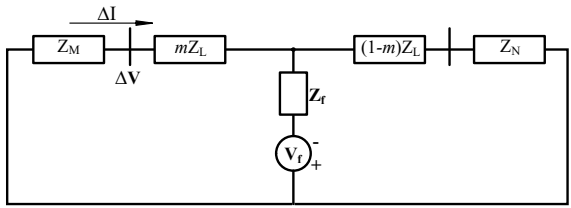}
    \caption{Pure Fault Network}
    \label{pure}
\end{figure}

Consider the simple network in Fig. \ref{fig_network}. When a fault occurs at F at a distance m from the relay at Bus M. The prefault and faulted networks can be obtained as shown in Fig. \ref{Prefault} and \ref{Fault}. By introducing a voltage source, V\(_{f}\), whose magnitude is equal to the pre-fault voltage at the fault point, and also by suppressing all sources (at M and N), a pure fault circuit can be obtained as shown in Fig. \ref{pure}. Thus the superimposed voltages and currents are the difference between the pre-fault and faulted voltages and currents respectively.

\begin{equation}
\Delta V = V_{fault} - V_{pre-fault} 
\label{Vpostfault}
\end{equation}
\begin{equation}
\Delta I = I_{fault} - I_{pre-fault}
\label{Ipostfault}
\end{equation}
where \(\Delta {V}\) and \(\Delta {I}\) represent the superimposed voltage and current respectively. It can also be shown that this applies to both the phase and sequence quantities \cite{9750756, benmouyal1999superimposed}. Considering ABC phase rotation, and using phase A currents as reference
\begin{equation}
\Delta I_{a} = \Delta I_{0}+\Delta I_{1}+\Delta I_{2}
\label{delta_Ia}
\end{equation}
Thus, from (\ref{Ipostfault}) total phase A current, is given by
\begin{equation}
I_{f} = (\Delta I_{0}+\Delta I_{1}+\Delta I_{2})+I_{pre_fault}
\label{If}
\end{equation}
where \(\Delta I_{0}, \Delta I_{1}\) and \(\Delta I_{2}\) represent the superimposed zero, positive and negative sequence currents respectively. Since the pre-fault currents are due to the load, it can be observed that the superimposed currents are devoid of the load currents. Similar analyses can be extended for superimposed voltages.

In the same vein, superimposed voltages have been defined as the ratio of the voltages to the currents
\begin{equation}
\Delta Z = \frac{V_{fault} - V_{pre-fault}}{I_{fault} - I_{pre-fault}} = \frac{\Delta V}{\Delta I}
\label{deltaZ}
\end{equation}
Sequence quantities of the impedance value can also be evaluated in a similar way.

\subsection{Applications in Microgrid Protection}
Since superimposed quantities offer a set of fault components, their magnitudes and angles during faults can be distinctive. They are therefore employed in diverse conventional and modern schemes as the measured quantities for fault detection. In this case, they can improve the sensitivity and selectivity of detection methods, as needed for microgrids. Also, they are useful for high speed detection.

In adaptive protection, superimposed sequence quantities are useful for computing adaptive fault currents based on the impact factor of different DGs in the network. The method proposed in \cite{muda2016superimposed} determines the microgrid operation using the breaker statuses, and determines an adaptive fault current based on the impact factor of various DGs present. The impact factor is a ratio that adjusts the fault current levels, making it adaptive to the contributory DGs. The factor, \(k\), is determined to be 
\begin{equation}
k = \frac{|\Delta I_{1f}| - |I_{1pre}|}{|\Delta I_{1f}|}
\label{impact}
\end{equation}
Since the fault current contribution of the grid, and different DG types vary, the fault components and ultimately the ratio obtained from \ref{impact} will vary. Thus the adaptive fault current, \(I_{f}\)  can be obtained as
\begin{equation}
|I_{f}| = (|1_{1f}|+|I_{2f}|+|I_{0f}|)\times k
\label{adaptive}
\end{equation}

Directional schemes have traditionally been employed in active distribution networks to deal with the challenges that arise from bi-directional flow of fault currents. It was determined in \cite{4218815} through analyses of different measured quantities that superimposed quantities provide the most accurate results for detection of fault direction. Thus, many schemes update existing directional protection with superimposed quantities.For instance, the pure fault sequence quantities are effectively used as polarizing quantities. 
Equation \ref{deltaZ2} shows a proposed improvement in \cite{7981375} to the traditional T32Q element used detecting the direction of unbalanced faults. 
\begin{equation}
\Delta Z_{2} = \frac{V_{2f} - V_{2pre}}{I_{2f} - I_{2pre}} = \frac{\Delta V_{2}}{\Delta I_{2}}
\label{deltaZ2}
\end{equation}
This approach eliminates the impact of control system of IBDG feedforward compensation in the traditional methods.

Other, proposed solutions for microgrid protetion such as the differential and distance protection schemes attempt to improve the applicability of these solutions to distribution networks using superimposed quantities. For instance \cite{7745916} employs quantities to improve the restraint and operate regions of a differential scheme to achieve more accurate operation when applied in distribution networks. In some cases, the pure fault components are used as starting criterion for conventional schemes, to prevent false trips.

\section{Incremental Negative Sequence Admittance}
This section describes a novel superimposed negative sequence admittance approach to detecting microgrid faults.

This element is an approach to improve the sensitivity of negative sequence directional element in the presence of IBDGs. A negative sequence approach is chosen since it these quantities are present in significant levels in unbalanced faults which form 95 percent of distribution system faults.

The element is evaluated by
\begin{equation}
\Delta Y_{2} = \frac{I_{2f} - I_{2pre}}{V_{2f} - V_{2pre}} = \frac{\Delta I_{2}}{\Delta V_{2}}
\label{deltaY2}
\end{equation}
where \(\Delta Y_{2}\) represents the superimposed negative sequence admittance. 

Compared to the impedance approach described in (\ref{deltaZ2}), the admittance method ensures that the magnitude evaluated provides a larger magnitude for for the forward fault detection. This improves the sensitivity of the directional scheme.  More so, the method provides more accurate responses for microgrids in grid-connected mode due to the negative sequence supply by the grid. Although only a small amount of negative sequence may be present at the inverter terminals, by using the superimposed approach, the incremental value provides a significant magnitude difference, compared to the non-faulted case. More so, the VDE-AR-N 4120 Technical Connection Rules adopted by the German grid code, provides a effective injection-based model for IBDGs to provide negative sequence reactive current support. The model and its practical application are extensively discussed in \cite{HADDADI2020106573}.

The threshold for the minimum \(\Delta Y_{2}\) setting for forward fault detection can be set to be greater than the minimum value observed during normal operation, \(Y_{set}\). This setting must take into consideration the unbalanced loads.

Analysis of the phase angle obtained from \(\Delta Y_{2}\), it can be deduced that, for an ideal source, and a purely inductive system
\begin{equation}
\arg \{\frac{\Delta I_2}{\Delta V_2}\} = \left\{ 
  \begin{array}{ c l }
    90^{\circ} & \quad \textrm{for forward fault}  \\
    -90^{\circ}                 & \quad \textrm{for reverse fault.}
  \end{array}
\right.
\end{equation}

However, considering the impedances of the protected line, the directional criterion can be modiified as shown in (\ref{forward}) and (\ref{reverse}).

Given that \(\phi\) is the compensation based on the line angle, the criterion can be interpreted as
\begin{equation}
0^{\circ}+ \phi < arg \{\Delta Y_2\} < 180^{\circ}+\phi
\label{forward}
\end{equation}
for forward faults, and

\begin{equation}
180^{\circ}+ \phi < arg  \{\Delta Y_2\} < 360^{\circ}+\phi 
\label{reverse}
\end{equation}
for reverse faults

\begin{figure}[!ht]
\centering
\includegraphics[clip, trim=0.25cm 3cm 10cm 0.25cm, width=0.48\textwidth]{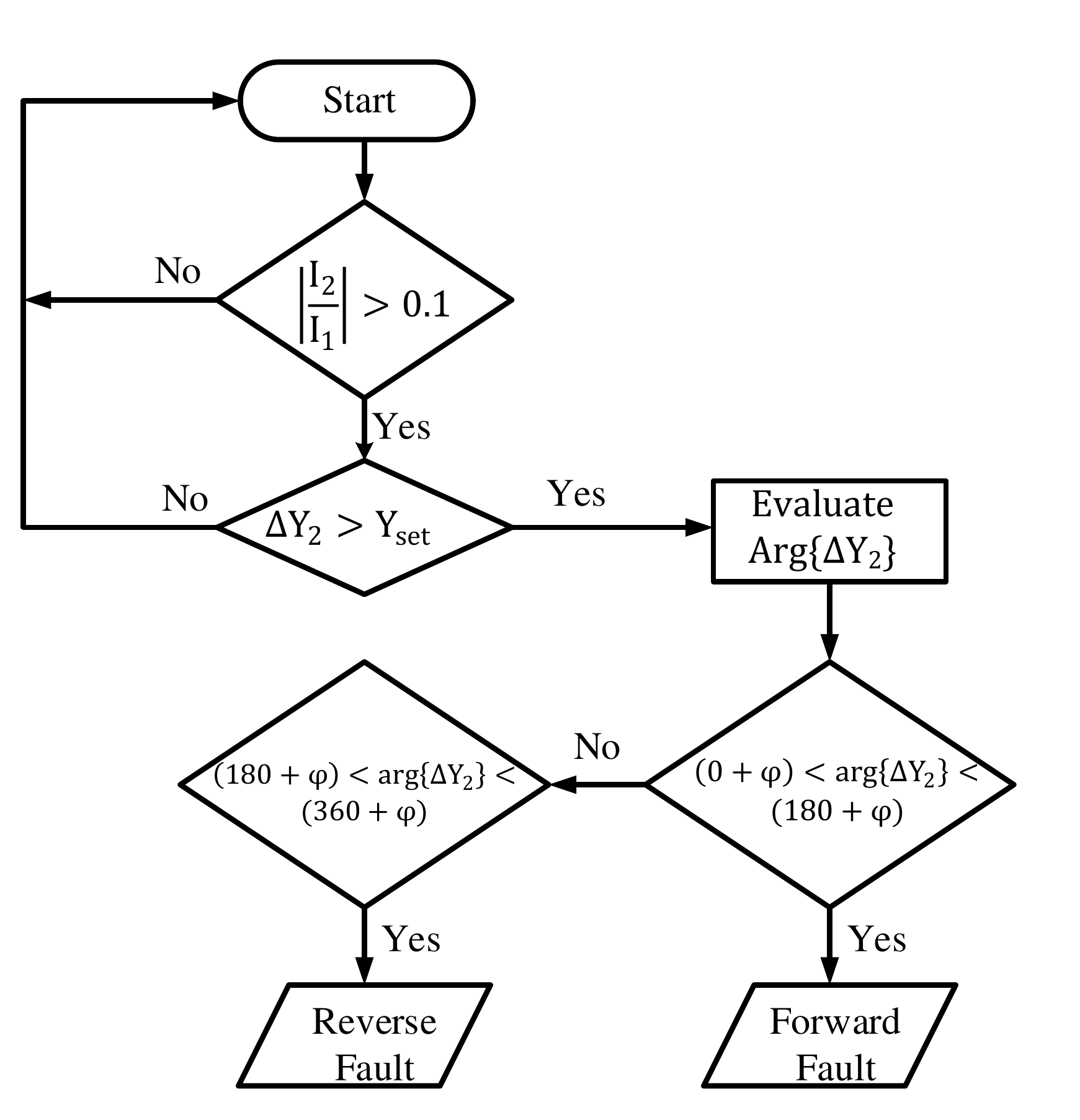}
\caption{Proposed Method}
\label{algo}
\end{figure}

As is used in other negative sequence based schemes, an important starting, \(\alpha\) condition is included to prevent false trips due to unbalanced load. This is a ratio factor of \(|I_{2}/I_{1}|\) is used and must be greater than 0.1 to initiate the fault detection algorithm.
A flowchart that summarizes the fault detection algorithm is presented in Fig. \ref{algo}.

\section{Test Results}
\begin{figure}[!t]
\centering
\includegraphics[clip, trim=2cm 10cm 4cm 0.5cm, width=0.6\textwidth]{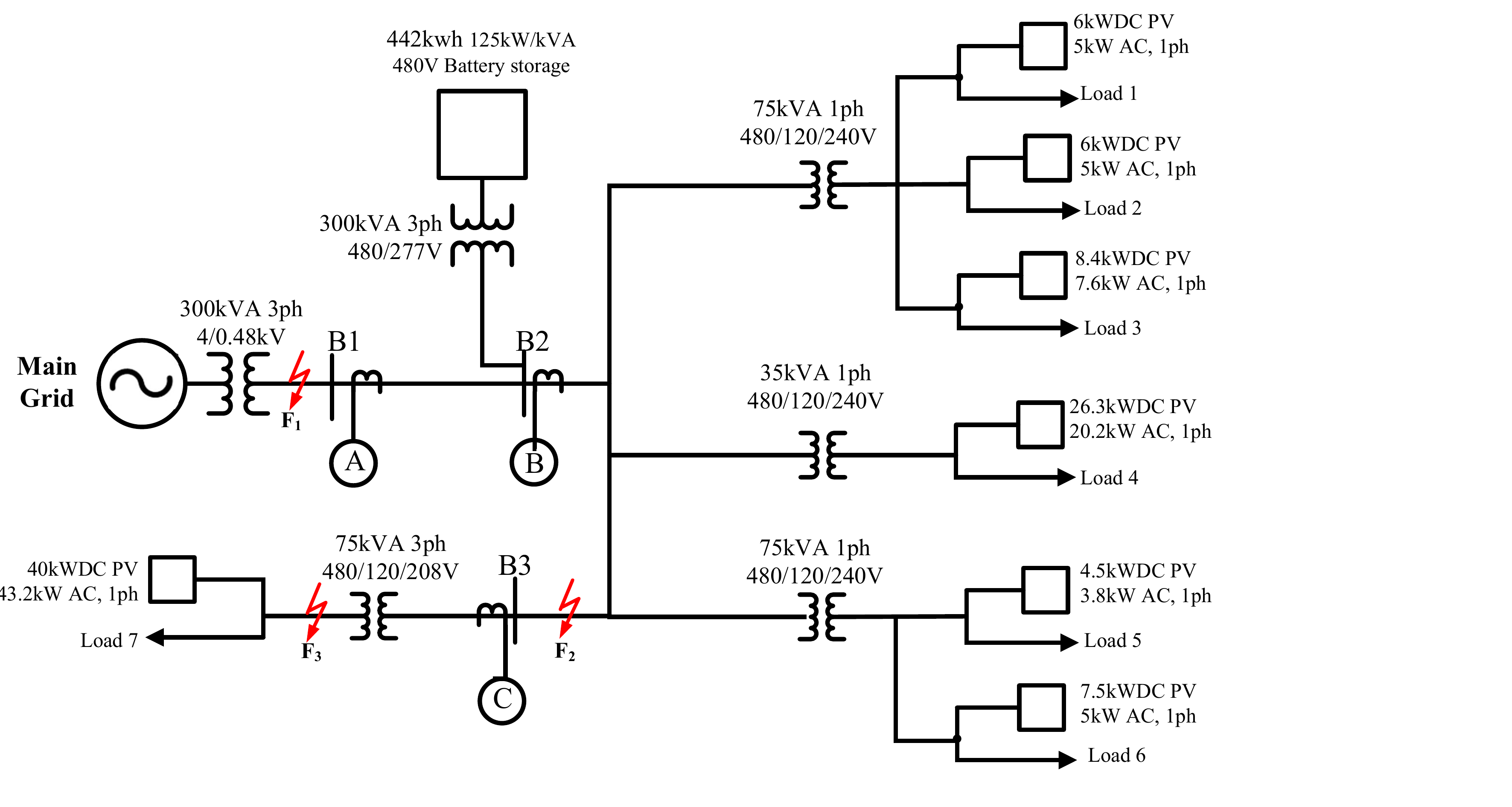}
\caption{Test Microgrid}
\label{MG}
\end{figure}

The proposed method was evaluated on a community microgrid, shown in Fig. \ref{MG}. The grid is connected at the substation to Bus 1 via a 300kVA 4/0.48kV transformer. A 442kWh battery energy storage unit consisting of consisting of two 125kW/kVA inverters is also connected to Bus 2 as a DG at the same substation. The loads consist of six single-phase connected and one three-phase connected commercial loads. The loads are served by 120/240V split-phase and 120/208 3P 4W service entrances. Each of these loads have individual rooftop PV with inverters and can each be islanded separately. 

Several simulation studies was carried out on the model to assess the performance of the protection system. Response of relays positioned at Bus 1, 2 and 3 were captured as shown in Figs \ref{F1_AC} to \ref{F3}. Tests were also carried out to determine the \(Y_{set}\) and \(\phi\) settings for the community microgrid. Protection engineers have guidance for default \(Z_{2}\) settings for traditional systems \cite{zimmerman2010fundamentals}. Although this could be adapted to obtain \(Y_{set}\), considerations must be made for applications involving active distribution networks with much more dynamic topology. This is reflected in a much wider variation of source impedances. Thus, from tests conducted on the system for normal operation, \(Y_{set}\) is set to be 10 for all relays A and B, and 5 for relay C. These tests consider the \(|\Delta Y_2|\) measured during faults are supplied the weakest sources \cite{7981375}. Also, considering the various impedances computed under these fault scenarios, \(\phi\), is set to a low value of 20\(^\circ\) for all the relays. Commercial relays set this value to be typically 20\(^\circ\) to 45\(^\circ\).

\subsection{Case A}
Case A shows the response of the relays 1 and 2 to B-C-G fault applied at 0.2s at the utility side, F\(_{1}\). Since Relays 1, 2 and 3 are set to be forward-looking towards the loads, a reverse direction decision is expected from all the relays. 

\begin{figure}[!t]
\centering
\includegraphics[width=0.48\textwidth]{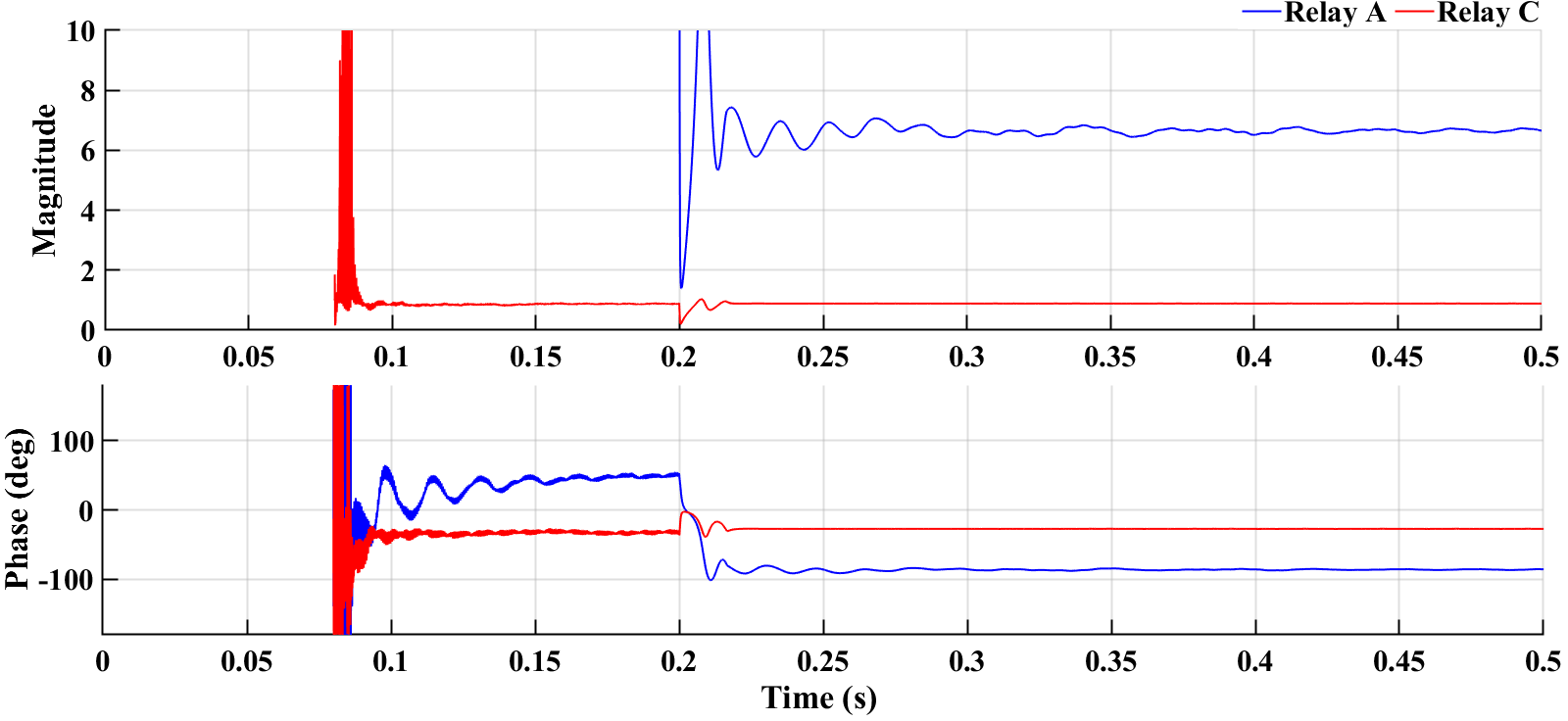}
\caption{\(\Delta Y_2\) response of Relay A and C for B-C-G fault at F\(_{1}\)}
\label{F1_AC}
\end{figure}  

As can be seen in Fig. \ref{F1_AC}, both relay A and C obtain an evaluation \(|\Delta Y_2|\), 6.89 and 0.88 respectively, that fall below the threshold of the normal operation once the fault is applied. Relay A with a smaller impedance loop results in a slightly larger magnitude but this is still below \(Y_set\) determined for forward fault operation. Similarly, arg\(\{\Delta Y_2\}\) evaluated for the relays are -98\(^\circ\) and -24\(^\circ\) respectively, thus for a \(\phi\) setting of 30\(^\circ\), a reverse fault decision is also determined by the phase angles. It is important that control systems are designed to to allow for a faster setting time for the relay response. Similarly, it must be noted that the readings recorded before the application of the fault at 0.2s, represent noise from the simulation and the control system. The algorithm is able to accurately avoid tripping as a result of the starting criterion included.

\begin{figure}[!t]
\centering
\includegraphics[width=0.48\textwidth]{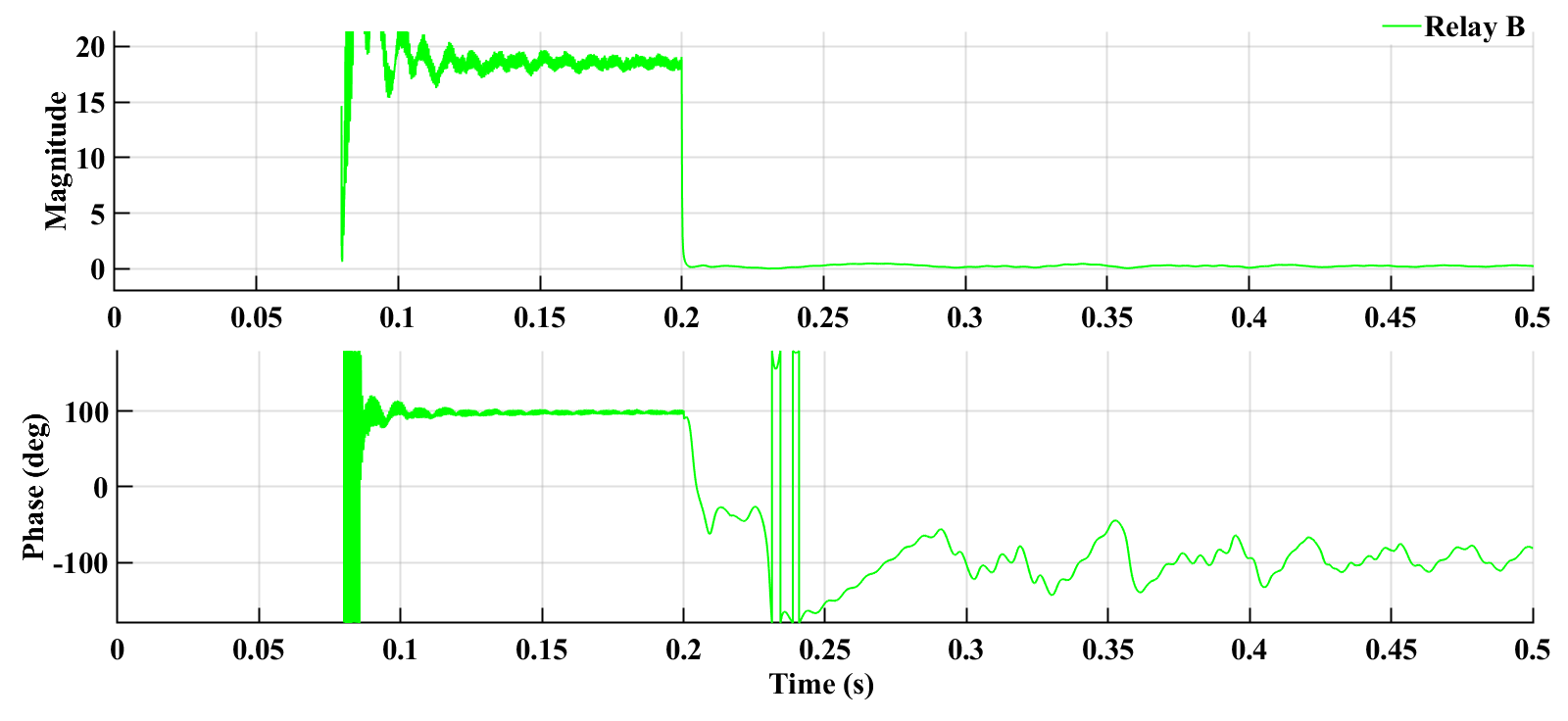}
\caption{\(\Delta Y_2\) response of Relay B for B-C-G fault at F\(_{1}\)}
\label{F1_B}
\end{figure}

Fig. \ref{F1_B} shows the response of Relay B. It is clear that the response of the single-phase inverters affect the stability of the arg\(\{\Delta Y_2\}\) obtained. However, the relay correctly determines the fault direction since the magnitude obtained is about 0.3. More so the oscillating phase is within the reverse zone.

\subsection{Case B}
Case B considers a fault, F\(_{2}\). This fault is on Load 7 feeder but is upstream of Relay C which is located at the load center. Thus, a reverse fault is the correct decision by Relay C. Both relays A and B which are upstream of the fault, must have a correct forward fault response. 

\begin{figure}[!t]
\centering
\includegraphics[width=0.48\textwidth]{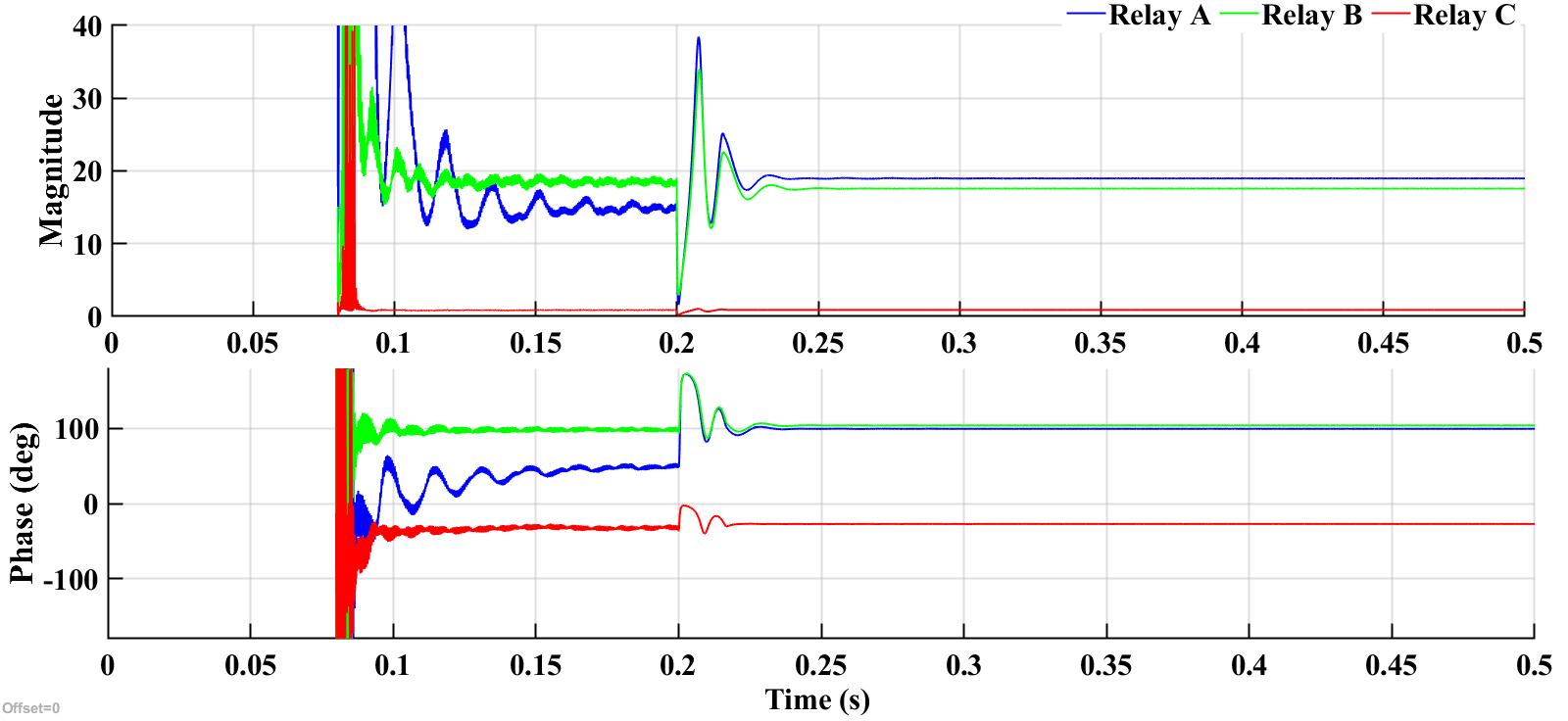}
\caption{\(\Delta Y_2\) response of relays for B-C-G fault at F\(_{2}\)}
\label{F2}
\end{figure}  

Fig. \ref{F2} demonstrates accurate discrimination of fault directions by Relays A and B by both magnitude and phase values. With the incidence of a forward fault, the \(|\Delta Y_2|\) values increase to about 19.8 and 17.9 respectively. Similarly, phase values at about 100\(^\circ\) are within the forward zone. Relay C, evaluates a correct reverse fault decision. 

An important observation is that, unlike in Case A (Fig. \ref{F1_B}), the grid-support allows for a faster settling time for the control action.

\subsection{Case C}
This case enables us to test the operation of load side relays, Relay C in this case, in the event of a forward fault, F\(_{3}\). An accurate operation of the relays achieves accurate discrimination between reverse faults F\(_{1}\),F\(_{2}\) and the reverse fault F\(_{3}\). 

\begin{figure}[!t]
\centering
\includegraphics[width=0.48\textwidth]{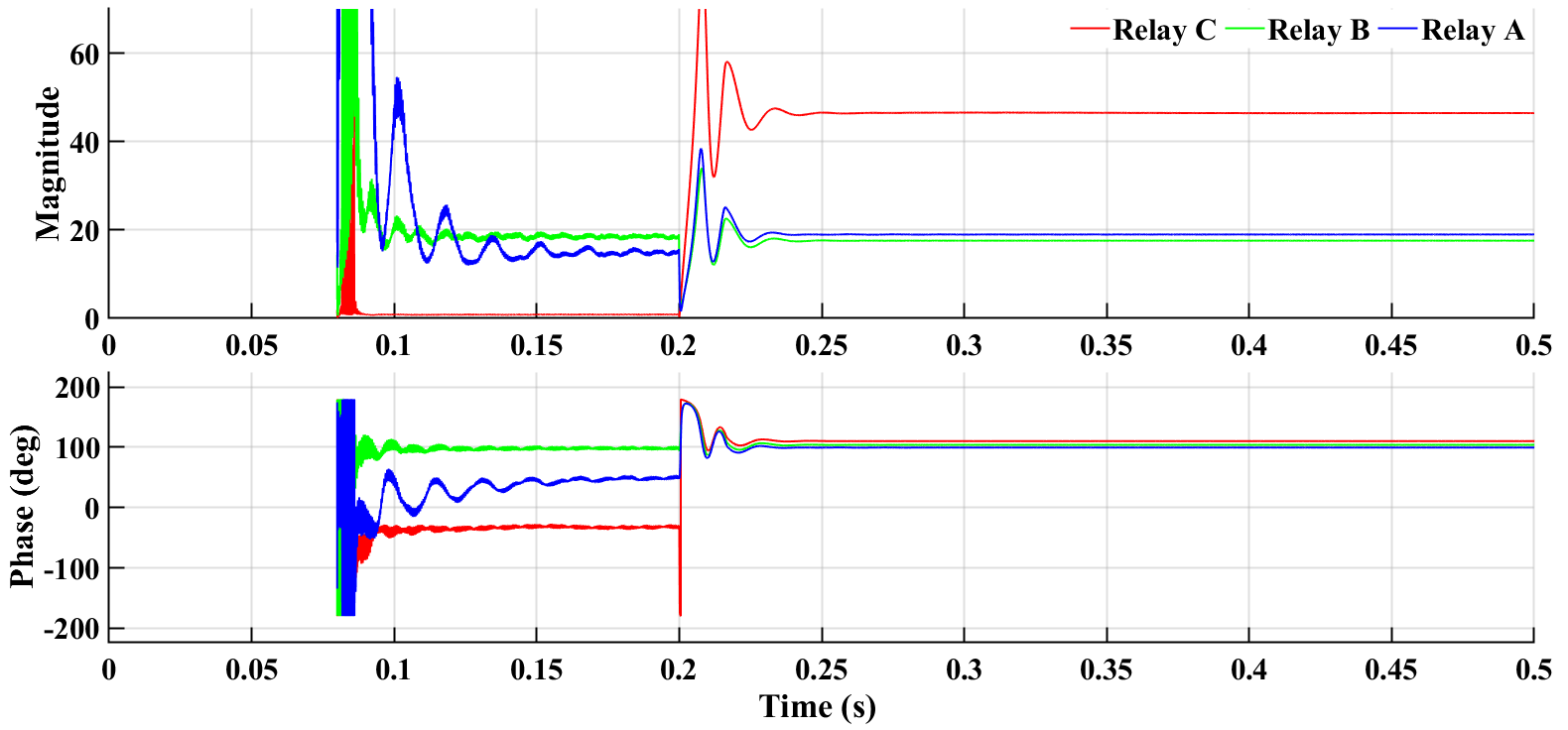}
\caption{\(\Delta Y_2\) response of Relay C for B-C-G fault at F\(_{3}\)}
\label{F3}
\end{figure}  
From Fig \ref{F3}, Relay C correctly identifies the forward fault. \(|\Delta Y_2|\) increases to about 48.9 whiles the arg\(\{\Delta Y_2\}\) of 105\(^\circ\) lies well within the forward zone. Relays A and B also, correctly determines the forward fault. It is important therefore, for coordination scheme to be included in this case.

\subsection{Case D}
Single-phase-to-ground faults were conducted in similar locations to monitor the response of the proposed method. These are the most occurring types of faults in distribution networks. Generally, it is expected that fault current in a single-phase-fault will be lower than other fault types in a similar location (depending on the transformer core type and winding connection, the positive and negative sequence impedance of the lines, fault resistance among other factors). The proposed negative sequence approach may therefore offer a better approach to detecting the lower fault current conditions.  

\begin{figure}[!t]
\centering
\includegraphics[width=0.48\textwidth]{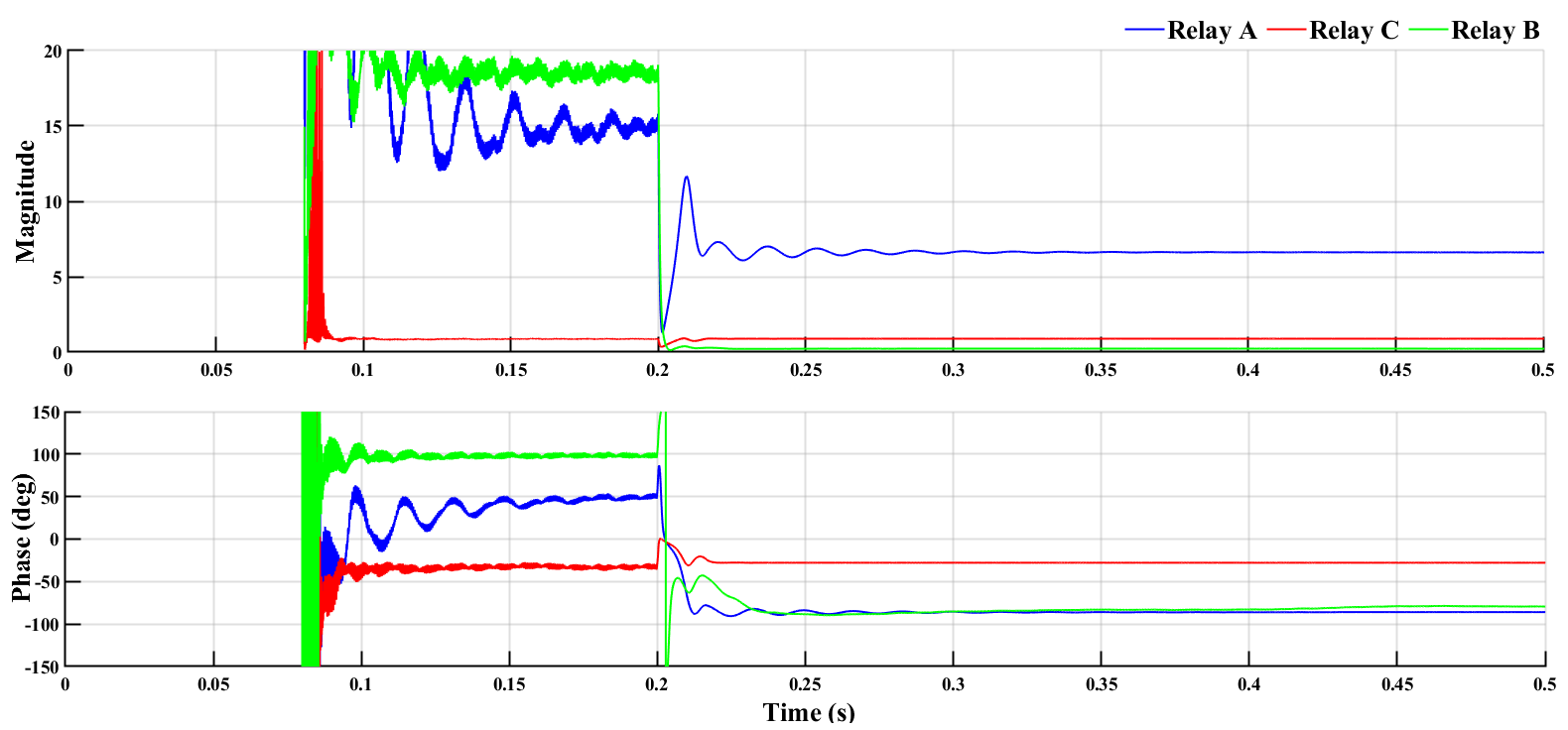}
\caption{\(\Delta Y_2\) response of Relay A,B,C for A-G fault at F\(_{1}\)}
\label{F1b}
\end{figure} 
Fig \ref{F1b}, demonstrates that, under A-G fault condition applied at 0.2s at F\(_{1}\), the \(\Delta Y_2\) is able to correctly detect the fault based on both the magnitude and the phase. It should be noted that, Relay at A evaluates a higher \(|\Delta Y_2|\) due to the nearness of the location of the fault to the relay A. Since the magnitude of negative sequence impedance obtained is closely related to the negative sequence impedance behind the relay, a lower superimposed negative sequence impedance seen by relay A because of the fault location (as compared to relays B and C), will translate to a higher magnitude of \(|\Delta Y_2|\). It is essential that forward direction thresholds for \(|\Delta Y_2|\) are determined conditions that lead to lowest  negative sequence impedances seen by relays under fault conditions.

\begin{figure}[!t]
\centering
\includegraphics[width=0.48\textwidth]{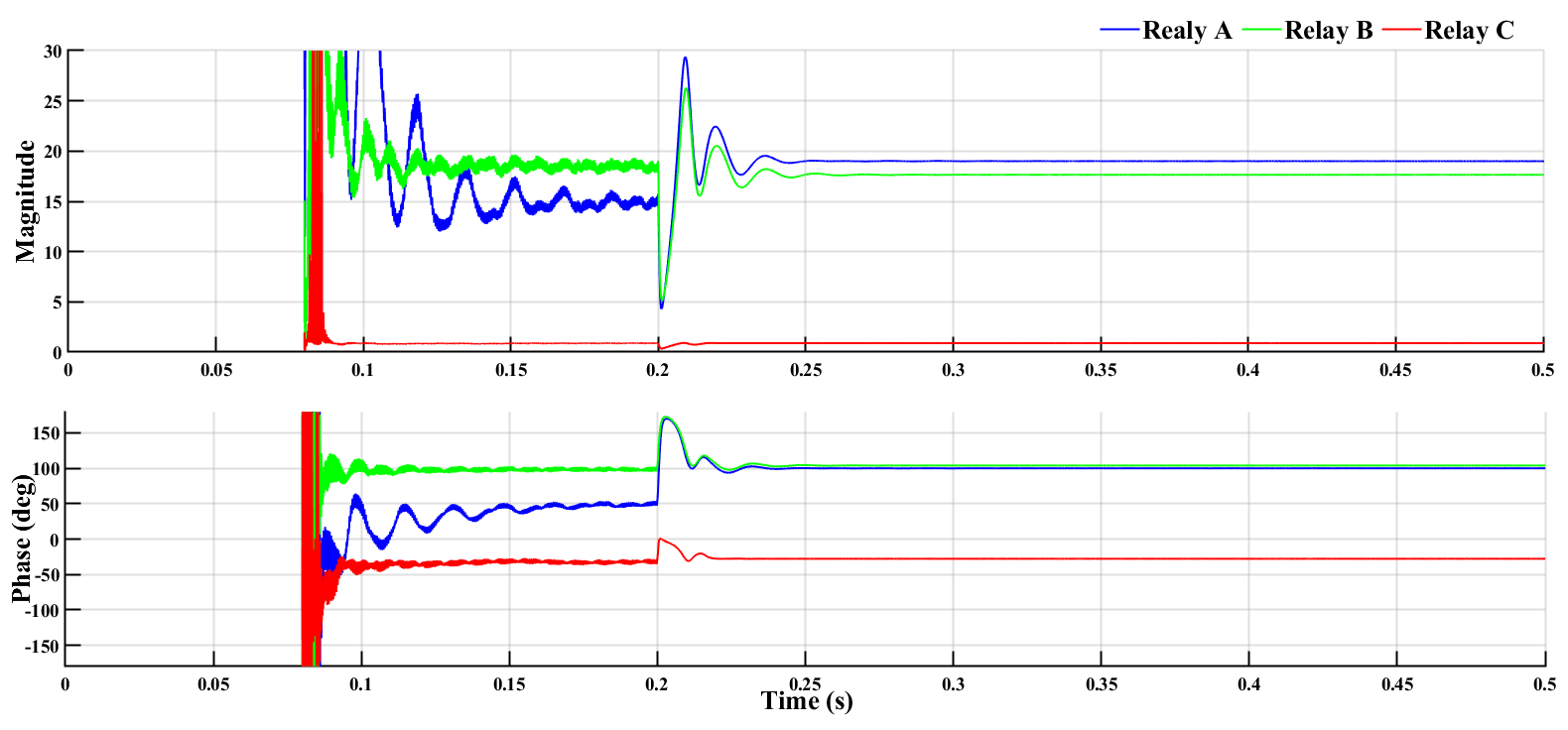}
\caption{\(\Delta Y_2\) response of Relay A,B,C for A-G fault at F\(_{2}\)}
\label{F2b}
\end{figure} 
A-G faults at F\(_{2}\) also produced correct detection response from the relays. It can be observed from \ref{F2b} that the effect of the three-phase inverter in the forward direction of relay C does not impact the superimposed sequence components evaluated by the relay and it correctly detects a reverse fault.

\begin{figure}[!t]
\centering
\includegraphics[width=0.48\textwidth]{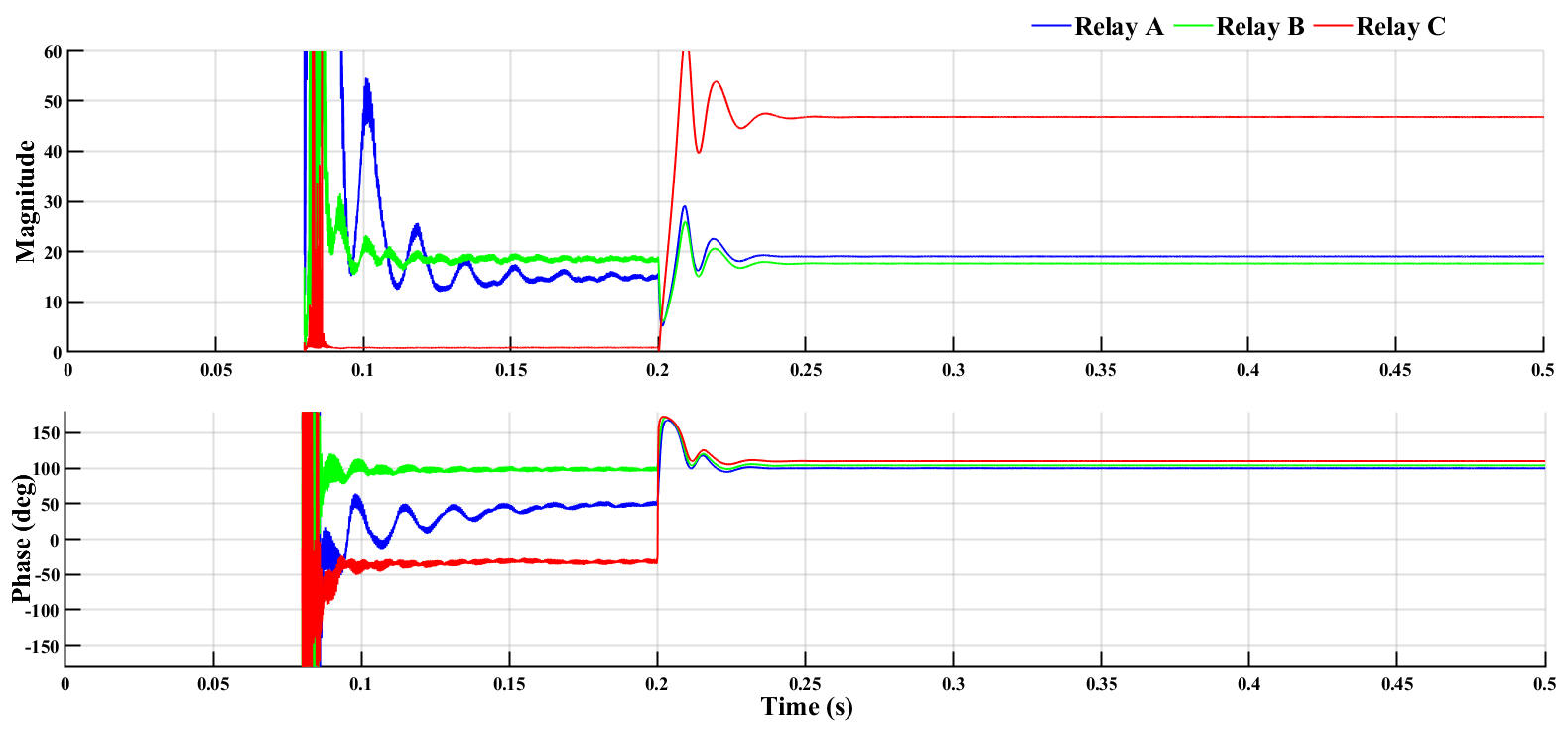}
\caption{\(\Delta Y_2\) response of Relay A,B,C for A-G fault at F\(_{3}\)}
\label{F3b}
\end{figure} 
For the fault at F\(_{3}\) all faults are expected to make a forward decision and this is accurately realized from \ref{F3b}. It can be seen that Relay at C evaluates a large magnitude. This demonstrates the advantage of the method in the grid-connected mode due to the significant contribution of negative sequence from the grid-side.

\section{Conclusion}
Solutions to microgrid protection challenges continue to evolve. It is however important that while improving on the various objectives of protection design, simplicity and economy in distribution protection are also maintained.

Directional protection offers an economical and simple method for dealing with the bidirectional issues of microgrid protection. When applied together with superimposed sequence quantities, the sensitivity of fault detection is improved, together with other protection objectives. The superimposed negative sequence admittance discussed in this paper successfully achieves this desired target.

It is important, however, for each system to be evaluated to determine proper threshold for the magnitude and phase of this element. While, negative sequence impedance meausurement may be adapted to determine the \(\Delta Y_2\) settings, impedance evaluation should take into consideration the variations due to inverter control action. More so, inverter design must ensure faster action to improve settling time for protection decisions.

\section*{Acknowledgement}
This material is based upon work supported by the U.S. Department of Energy’s Office of Energy Efficiency and Renewable Energy (EERE) under the Solar Energy Technologies Office Award Number DE-EE0002243-2144.

The views expressed herein do not necessarily represent the views of the U.S. Department of Energy or the United States Government.







\bibliographystyle{IEEEtran}


\bibliography{Reference}

\section*{Biographies}
\vskip -4\baselineskip plus -1fil
\begin{IEEEbiographynophoto}{Kwasi Opoku}
received a BSc in Electrical/Electronic Engineering from the Kwame Nkrumah University of Science and Technology, Kumasi, Ghana, in 2011. He is currently pursuing a PhD in Electrical Engineering at the University of Central Florida, Orlando, Florida. Before starting his PhD, he worked as a Protection Application Engineer at Schweitzer Engineering Laboratories Inc. and then as a Service Application Engineer with GE Grid Solutions, both in Ghana. His research interests include fault detection methods in inverter-interfaced microgrids, power system modeling and analysis.
\end{IEEEbiographynophoto}
\vskip -2\baselineskip plus -1fil
\begin{IEEEbiographynophoto}{Subash Pokharel}
received a B.E. degree in Electrical Engineering from Tribhuvan University, Nepal, in 2016. He is currently pursuing the Ph.D. degree in Electrical Engineering at the University of Central Florida, Orlando, FL. His research interest includes modeling, analysis, and optimization of magnetic power control devices along with power system analysis and protection.
\end{IEEEbiographynophoto}
\vskip -2\baselineskip plus -1fil
\begin{IEEEbiographynophoto}{Aleksandar Dimitrovski}
is an Associate Professor at the University of Central Florida, Orlando. Before joining UCF, he was a Chief Technical Scientist at the Oak Ridge National Laboratory and a Joint Faculty at the University of Tennessee, Knoxville. In the past, he had been with Schweitzer Engineering Laboratories, and Washington State University, Pullman. He received his B.Sc. and Ph.D. degrees in Electrical Engineering with emphasis in power, and M.Sc. degree in Applied Computer Sciences in Europe. His area
of interest has been focused on modeling, analysis, protection, and control of uncertain power systems and, recently, on hybrid magnetic-electronic power control devices.
\end{IEEEbiographynophoto}

%

\end{document}